\documentclass[twocolumn,a4paper,10pt]{article}
\usepackage{amsmath,amsfonts}
\usepackage{epsfig}
\usepackage{times}
\usepackage{epsfig}
\usepackage[small,hang]{caption}
\usepackage{graphicx}

\usepackage{subcaption}
\usepackage{natbib}
\usepackage{url}


\title{\vspace*{-12mm}
\LARGE \sc \textbf{  
Impact of Wind Direction on Flow Over a Realistic Urban Area: A Large-Eddy Simulation Study
}}

\author{ \Large \bf \textit{ 
I. Rodríguez $^{1}$, J.M. Duró$^{1}$, E. Mestres $^{1}$, M. Teng$^{2}$ and O. Lehmkuhl$^{2}$}  \\ \\
\bf  $^{1}$ \textit{ Universitat Politècnica de Catalunya} \\
\bf  $^{2}$ \textit{ Barcelona Supercomputing Center} \\
{\it ivette.rodriguez@upc.edu}
}
\date{}

\begin{document}
\maketitle
\thispagestyle{empty}

\section*{Abstract}
We conducted high-resolution large-eddy simulations over a real urban district in Barcelona to examine the impact of wind direction on near-ground flow. The computational mesh resolves over 500 million degrees of freedom, with a spatial resolution on the order of 1 m at pedestrian level. This allows a detailed analysis of mean velocity and turbulence patterns within the canopy layer. Although instantaneous flow fields differ significantly between cases, double-averaged profiles of velocity and turbulence intensity remain remarkably consistent across all wind directions. The results reveal a shear-driven mixing layer below the average building height and turbulence maxima near the tallest buildings, highlighting the influence of urban morphology on the development of flow and turbulence.

\section{Introduction}

Urban environments are characterized by complex interactions between buildings and atmospheric flows, leading to highly unsteady and three-dimensional wind patterns. Understanding the effect of wind direction on urban airflow is crucial for a wide range of applications, including air quality assessment, pedestrian comfort,  urban resilience to extreme weather events, among others.

Due to the irregular geometry of urban settings and the turbulent nature of the flow, numerical simulations have become essential tools for studying urban aerodynamics.   However, the complex phenomenology and geometry present make high-fidelity simulations usually very costly (e.g. \cite{hertwig2017les} ). In this sense, urban canopy models have been used to capture the main momentum exchange process at affordable cost. 
The urban canopy model is based on parameterizations suitable for large-scale roughness elements \citep{coceal2004canopy}. However, its application to real urban canopies presents challenges, particularly when considering the effect of wind direction. Despite its importance, there is limited information in the literature on studies that examine the influence of wind direction using realistic urban geometries. A deeper understanding of how wind flow direction impacts turbulence dynamics within real urban areas is crucial for improving urban airflow modeling and its related applications.

In this study, we investigate the impact of wind direction on airflow dynamics within an urban environment using large-eddy simulations (LES). % Due to the complexity of the flow most of the studies of urban flows performed so far has been done using  RANS (e.g. \cite{Toparlar2015,Brozovsky2021,Antoniou2019}) 
%or using large-eddy simulations (LES) in scaled down geometries. Only a few works reported in the literature use LES in complex urban settings (e.g. \cite{Lee2019,Yan2022}) showcasing the importance of high-fidelity simulations to study the complex turbulent interaction between buildings and the atmospheric boundary layer (ABL). In most of the studies reported so far, the effect of wind direction over the urban geometry has received less attention. \cite{xie2011wind}  presented simulations examining how varying wind directions influence airflow over arrays of cubic obstacles, representative of urban canopies. The findings highlighted that even slight deviations in wind direction can significantly alter flow patterns, which has important implications for urban planning and pollutant dispersion modeling.
Our analysis focuses on how variations in the approaching wind direction over a realistic urban geometry influence key flow characteristics. We consider different wind directions to assess their effect on flow separation, wake dynamics, as well as turbulence statistics. The results of this study will contribute to a deeper understanding of urban wind behavior and ultimately provide insights that can inform urban planning, pollutant dispersion modeling, and building design strategies.

\section{Mathematical and numerical model}
The filtered incompressible Navier-Stokes equations have been numerically solved using SOD2D (Spectral high-Order coDe for solving partial Differential equations), a low-dissipation spectral element method (SEM) code \citep{GASPARINO2024109067}.

SOD2D is based on a spectral-element version of Galerkin's finite element method continuous model. Projection stabilization prevents numerical oscillations caused by dominant convection while introducing minimal numerical dissipation. The aliasing effects from reduced-order SEM integration of convective terms are mitigated using a skew-symmetric splitting approach \citep{kennedy2008reduced}.

For temporal discretization, a BDF-EXT3 high-order operator splitting method is used to solve the velocity-pressure coupling \citep{Karniadakis1991}.

SOD2D is designed to efficiently leverage modern computational power, particularly GPUs, which play a central role in high-performance computing. It is capable of running on both GPU and CPU architectures. %The code is written in Fortran and employs MPI and OpenACC to enable parallelism at both coarse- and fine-grained levels. %The mesh is partitioned using the GEMPA library, while HDF5 is used for I/O, both of which are efficient and widely tested libraries for HPC.

%\subsection{Double-Averaging turbulent statistics}

In urban canopy flows, the high degree of spatial heterogeneity introduced by buildings and streets necessitates specialized averaging techniques to meaningfully analyze turbulence statistics. One widely adopted approach is the {double-averaging method} \citep{Raupach1982}, which extends the traditional Reynolds decomposition by including a spatial average in addition to the temporal one. The instantaneous flow variable $\phi(\mathbf{x}, t)$, such as a velocity component, is first decomposed as
$
\phi(\mathbf{x}, t) = \overline{\phi}(\mathbf{x}) + \phi'(\mathbf{x}, t),
$  where $\overline{\phi}$ is the temporal average at point $\mathbf{x}$, and $\phi'$ is the temporal fluctuation. A spatial average, denoted by angle brackets $\langle \cdot \rangle$, is then applied to the temporally averaged field, yielding
$ 
\phi(\mathbf{x}, t) = \langle \overline{\phi} \rangle + \phi''(\mathbf{x}) + \phi'(\mathbf{x}, t),
$  where $\langle \overline{\phi} \rangle$ is the combined spatial and temporal average, $\phi'' = \overline{\phi} - \langle \overline{\phi} \rangle$ is the spatial deviation from the mean, and $\phi'$ remains the temporal fluctuation.

\section{Definition of the cases}

\begin{table}[]
\centering
\caption{Geometric properties of the Barcelona domain and porosity metrics for different wind directions}
\begin{tabular}{l c c c c}
\hline
\text{Wind direction} & \textbf{0°} & \textbf{67.5°} & \textbf{90°} & \textbf{180°} \\
\hline
\multicolumn{5}{l}{\textbf{}} \\
Topography & \multicolumn{4}{c}{\includegraphics[width=3.5cm]{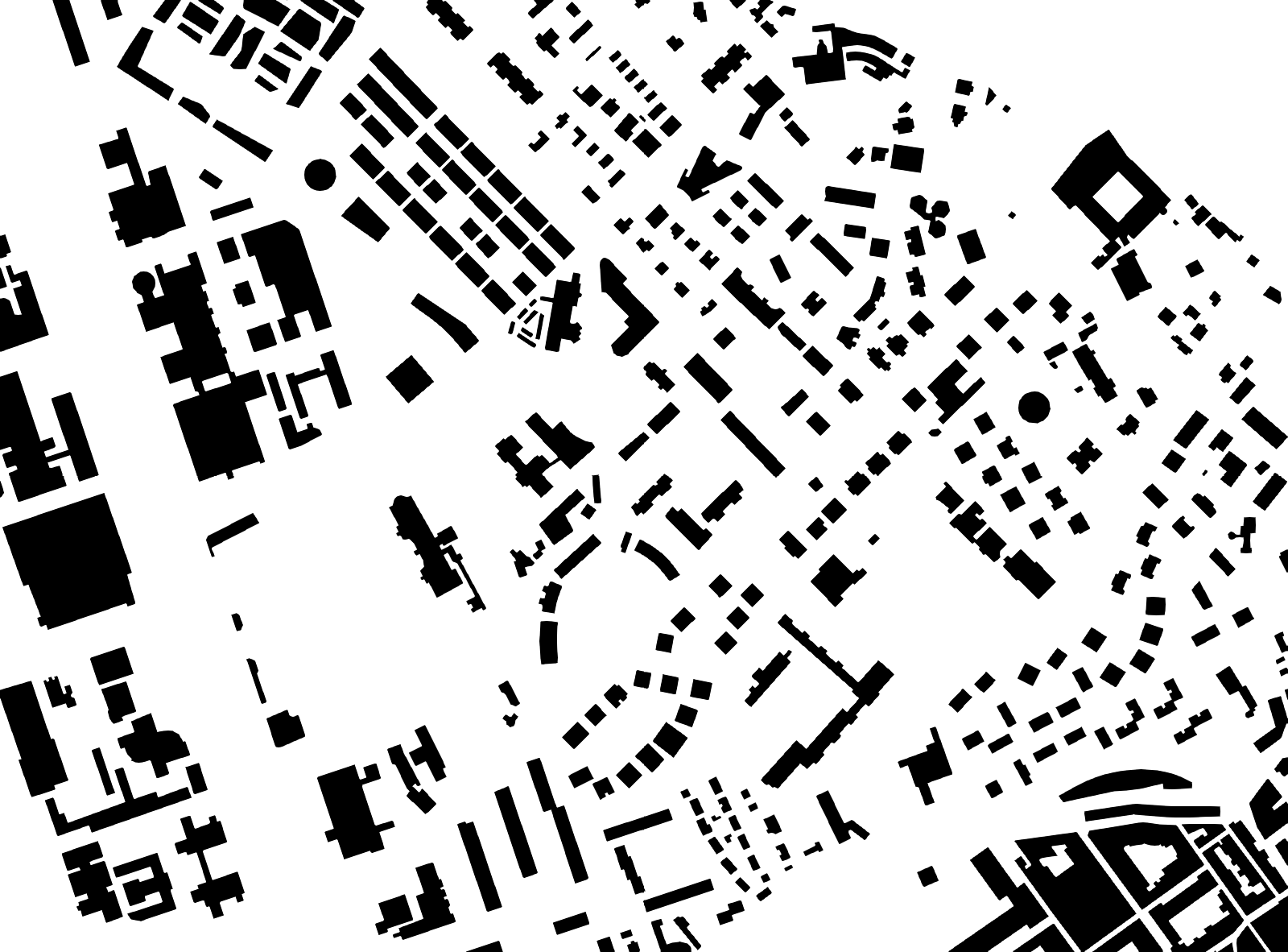}} \\
$H_{avg}$ (m) & \multicolumn{4}{c}{19.27} \\
$H_{std}$ (m) & \multicolumn{4}{c}{8.49} \\
$H_{max}$ (m) & \multicolumn{4}{c}{67.29} \\
$H_{min}$ (m) & \multicolumn{4}{c}{2.79} \\
\hline
DOF $\times10^6$ & 519 & 476& 487 & 480\\
$\lambda_p$ & 0.24 & 0.24 & 0.24 & 0.24 \\
$\lambda_f$ & 0.022 & 0.019 & 0.023 & 0.022 \\
\hline
\end{tabular}
\label{tab:barcelona_wind_directions}
\end{table}

\begin{figure*}
  \centering
  \begin{subfigure}[b]{0.35\textwidth}
    \includegraphics[width=\textwidth]{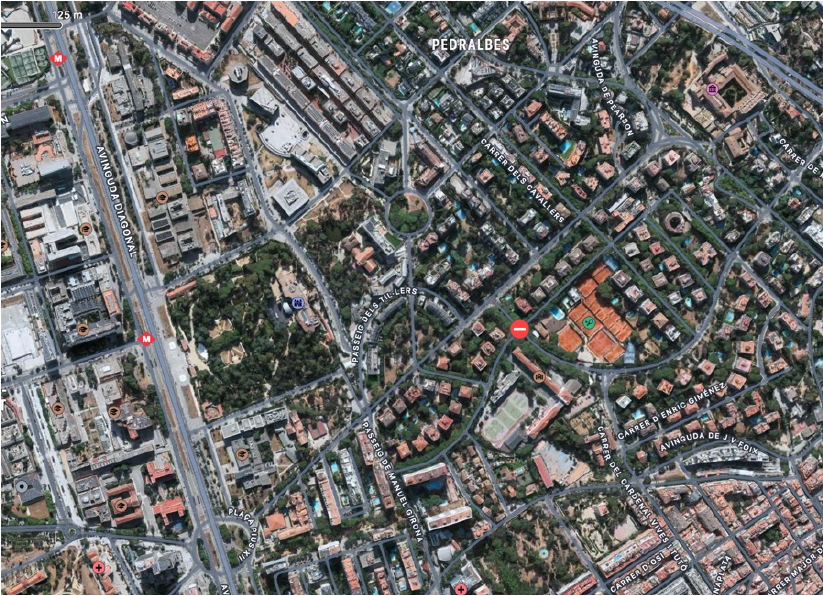}
    \caption{}
    \label{fig:line1_u}
  \end{subfigure}
  \hfill
  \begin{subfigure}[b]{0.49\textwidth}
    \includegraphics[width=\textwidth]{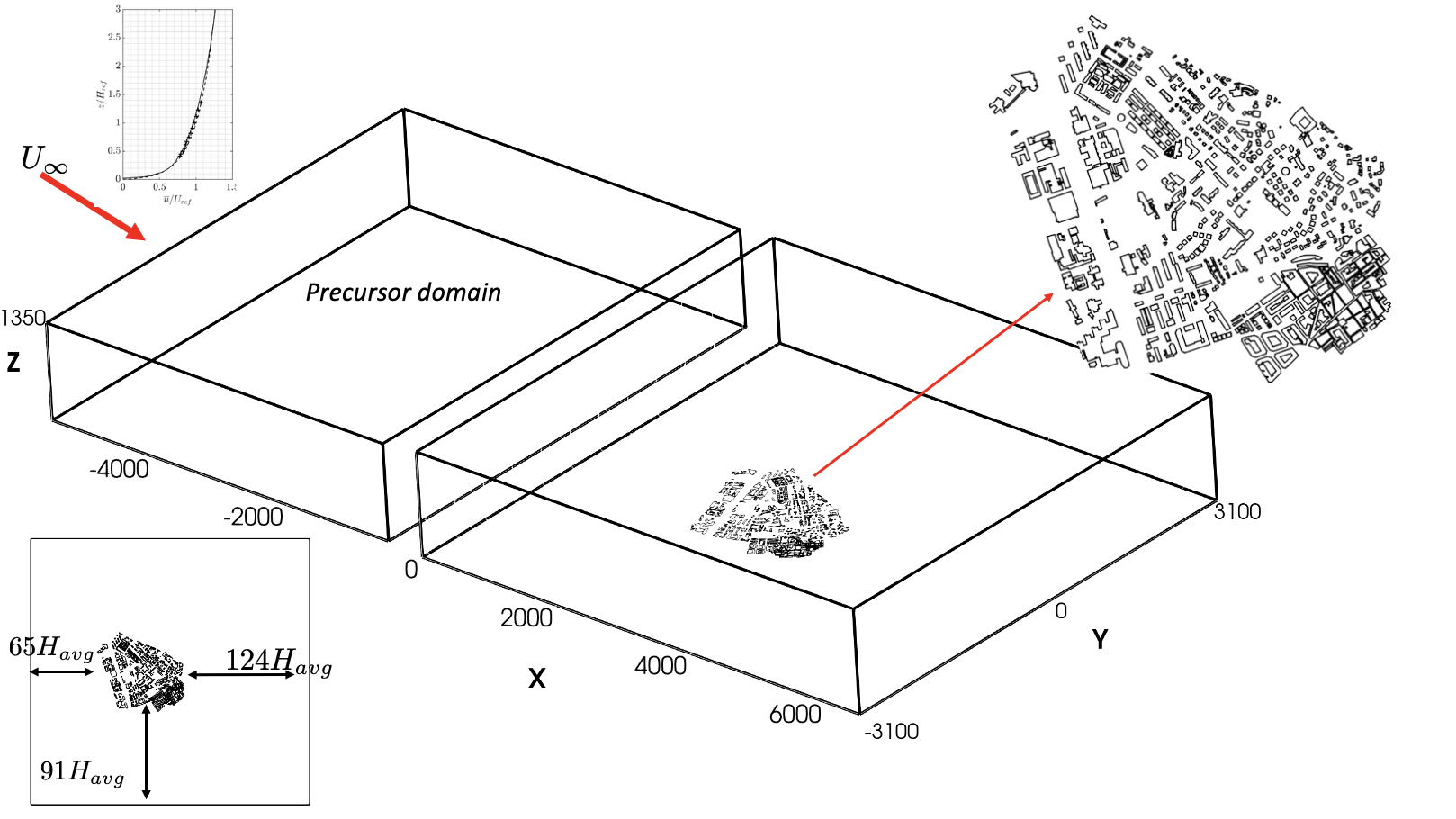}
    \caption{}
    \label{fig:line1_urms}
  \end{subfigure}
  \caption{(a) Map of Barcelona including the zone under study. (b) computational domain including the online precursor  domain }
  \label{fig:domain}
\end{figure*} 
To evaluate the impact of the wind direction, 
the study focuses on a neighborhood in Barcelona covering an area of approximately 1.7 $\times$ 1.9 km$^2$, characterized by significant morphological complexity (see Figure~\ref{fig:domain}a). The tallest structure within the domain reaches up to 67 m. Four distinct wind inflow directions are considered: $\phi = 0^\circ$, $67.5^\circ$, $90^\circ$, and $180^\circ$.

Table~\ref{tab:barcelona_wind_directions} summarizes the main morphological descriptors of the domain, including the mean building height ($H_{avg}$), the standard deviation of building heights ($H_{std}$), and the minimum and maximum heights ($H_{min}$ and $H_{max}$). The table also reports the plan area fraction $\lambda_p = A_p/A_{total}$ and the frontal area fraction $\lambda_f = A_f/A_{total}$, where $A_p$ and $A_f$ denote the total plan and frontal areas of the buildings, respectively, and $A_{total}$ corresponds to the total plan area of the simulation domain.

In Fig. ~\ref{fig:domain}, a street view of the zone of interest is shown together with the computational domain used for the simulations. While the example corresponds to wind direction $\Phi = 0^\circ$, the domain setup is representative of all simulated wind directions.
 To consider the different wind directions, the city is rotated accordingly, while the inlet is kept from left to right. Following previous results in  \cite{Teng2025}, where we computed a city-like urban scenario, distances from the city to the boundaries have been set. To accurately resolve the flow around complex urban geometries, unstructured hexahedral meshes were constructed with fourth-order polynomial basis functions used within each element. The mesh is constructed using nested refinement regions to ensure increased resolution near the ground, achieving a spatial resolution of approximately 1m at pedestrian level.  As a result, meshes in the order of 500 million grid points are solved for each case (see also Table~\ref{tab:barcelona_wind_directions} ).  To reduce computational cost while ensuring a reliable initialization, each simulation is first advanced using a second-order mesh with the same number of elements as the target high-order configuration. This low-order setup allows the flow to evolve efficiently until a statistically steady state is reached. Once transients have been washed out and equilibrium is established, the flow field is interpolated onto the fourth-order mesh, from which the high-fidelity simulation is re-started.
Following this initialization procedure, approximately \textbf{36\,$T$} (non-dimensional time units) are required for the flow to reach a new statistically steady state on the high-order mesh. Here, \textbf{$T$} is defined as \( T = H_{\text{avg}} / u^* \), where \( H_{\text{avg}} \) is the average building height and \( u^* \) the friction velocity. After this spin-up period, the simulation is continued for an additional \textbf{230\,$T$} to collect statistically converged flow statistics.

To generate realistic turbulent inflow for the urban domain, a precursor simulation runs concurrently with the main simulation, forming a coupled dual-domain system. The precursor domain, driven by a constant pressure gradient, develops a statistically stationary atmospheric boundary layer (ABL) under periodic conditions in the streamwise ($x$) and spanwise ($y$) directions. At each time step, the velocity field at the precursor outlet is directly mapped to the main domain's inlet, ensuring temporally coherent inflow.
In this setup, a reference wind speed of approximately 6.1 m/s is maintained at 100 m height, and the ABL is configured to reflect a highly rough surface with a roughness length ($z_0$) of 1.53 m, corresponding to a friction velocity ($u_*$) of 0.596 m/s. Free-slip conditions are applied at the top and lateral boundaries of both domains, and a fixed static pressure is imposed at the outlet of the main domain. Inside the urban canopy, the flow is highly three-dimensional, with separation and recirculation occurring near buildings. These complexities make equilibrium-based wall models unsuitable for vertical surfaces. As a result, no-slip conditions are enforced on building walls, while a wall model is used only at the ground to balance accuracy and cost.

\section{Results}

\subsection{Validation of the results}
For validation purposes, two cases of increasing complexity have been considered.
The first case studied is the wind tunnel experiment conducted by \cite{Brown2001}. The experiment involved a neutral atmospheric boundary layer (ABL) approaching a scaled \(1:200\) matrix of \(7 \times 7\) cubes, each with dimensions \(L = W = H = 0.15\) m (where \(L\), \(W\), and \(H\) represent length, width, and height, respectively). In the numerical simulations, the full-scale geometry is modeled, with \(H\) corresponding to 30 m.  

\begin{figure*}[h]
  \centering
  \begin{subfigure}[b]{0.49\textwidth}
    \includegraphics[width=\textwidth]{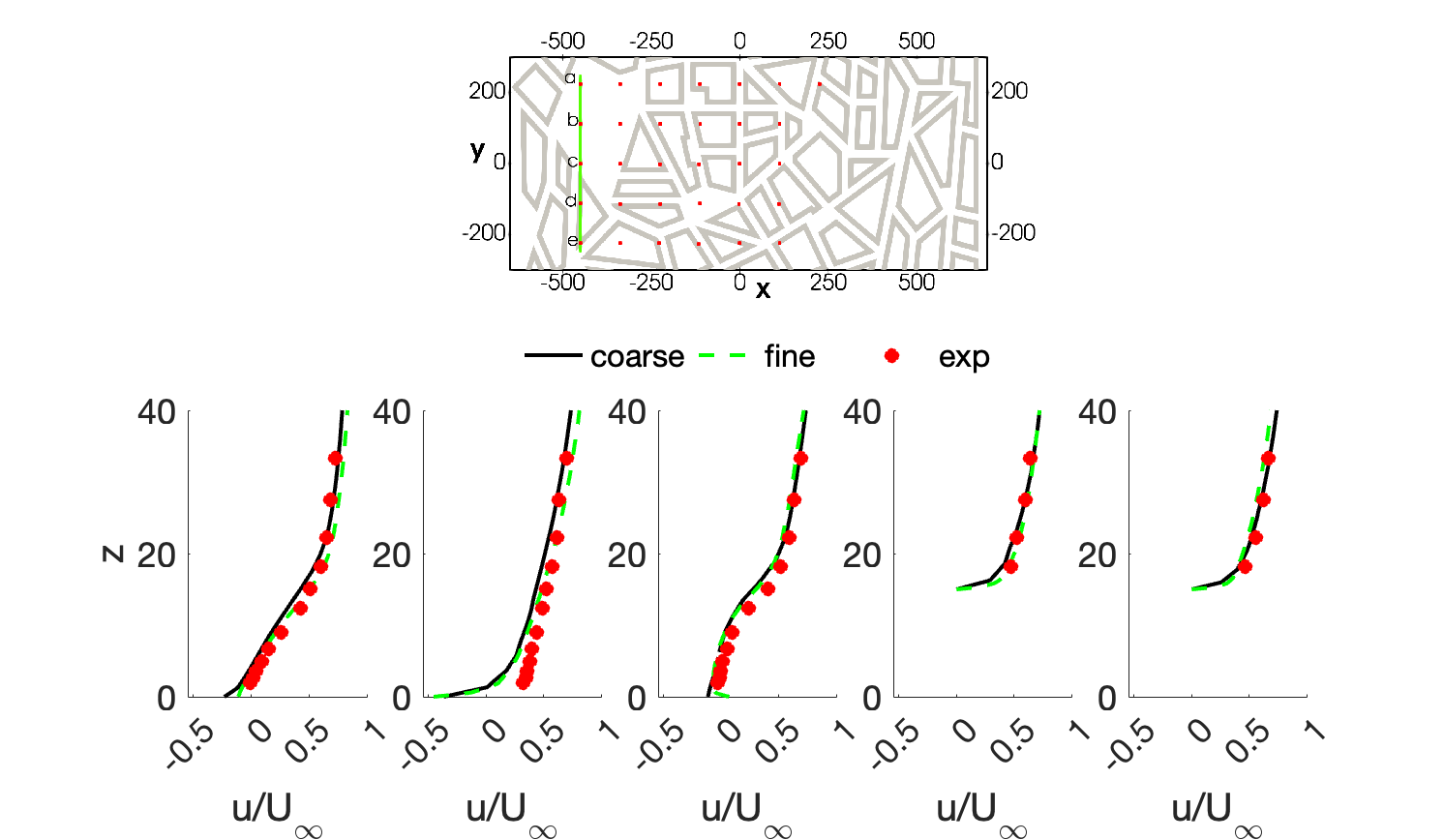}
    \caption{}
    \label{fig:line1_u}
  \end{subfigure}
  \hfill
  \begin{subfigure}[b]{0.49\textwidth}
    \includegraphics[width=\textwidth]{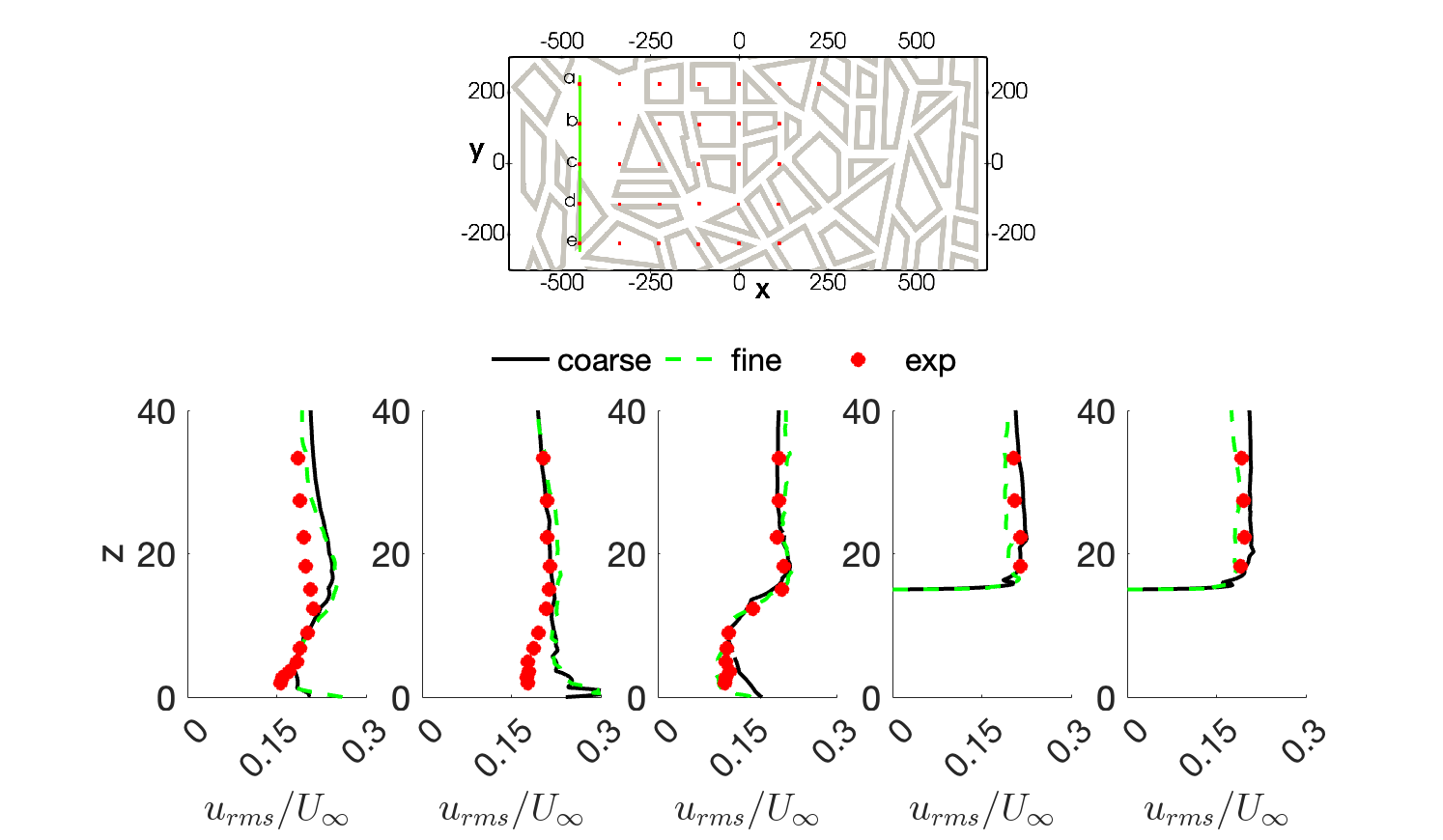}
    \caption{}
    \label{fig:line1_urms}
  \end{subfigure}
  \caption{Comparison between the different meshes and experimental data from \cite{leitl2024}. (a)Vertical profiles of the streamwise velocity at different stations. (b)Root-mean-square (rms) values of the fluctuations at different stations.}
  \label{fig:F3}
\end{figure*}

The second case simulated is a semi-idealized city ("Michel-Stadt"; CEDVAL-LES database, case reference: BL3-3) (\cite{leitl2024}).  
The modeled city area measures \( 1,320 \,\text{m} \times 830 \,\text{m} \), with all buildings featuring flat roofs to simplify wind flow interactions. 
For simplicity, only the results for this case are presented here. For a more details about the validation the reader is referred to \cite{Teng2025}. To compute this case, 40 flow-throughs are collected to gather statistical data. In Figure \ref{fig:F3} results obtained with the different levels of refinement in comparison against results from wind tunnel are presented. Both streamwise velocity and root-mean-square (rms) values of its fluctuations are pretty well predicted with the different meshes.

\subsection{Wind effects at pedestrian level}

\begin{figure*}[h]% "t" para top de la página
  \centering
  \begin{subfigure}[b]{0.32\textwidth}
    \includegraphics[width=\textwidth]{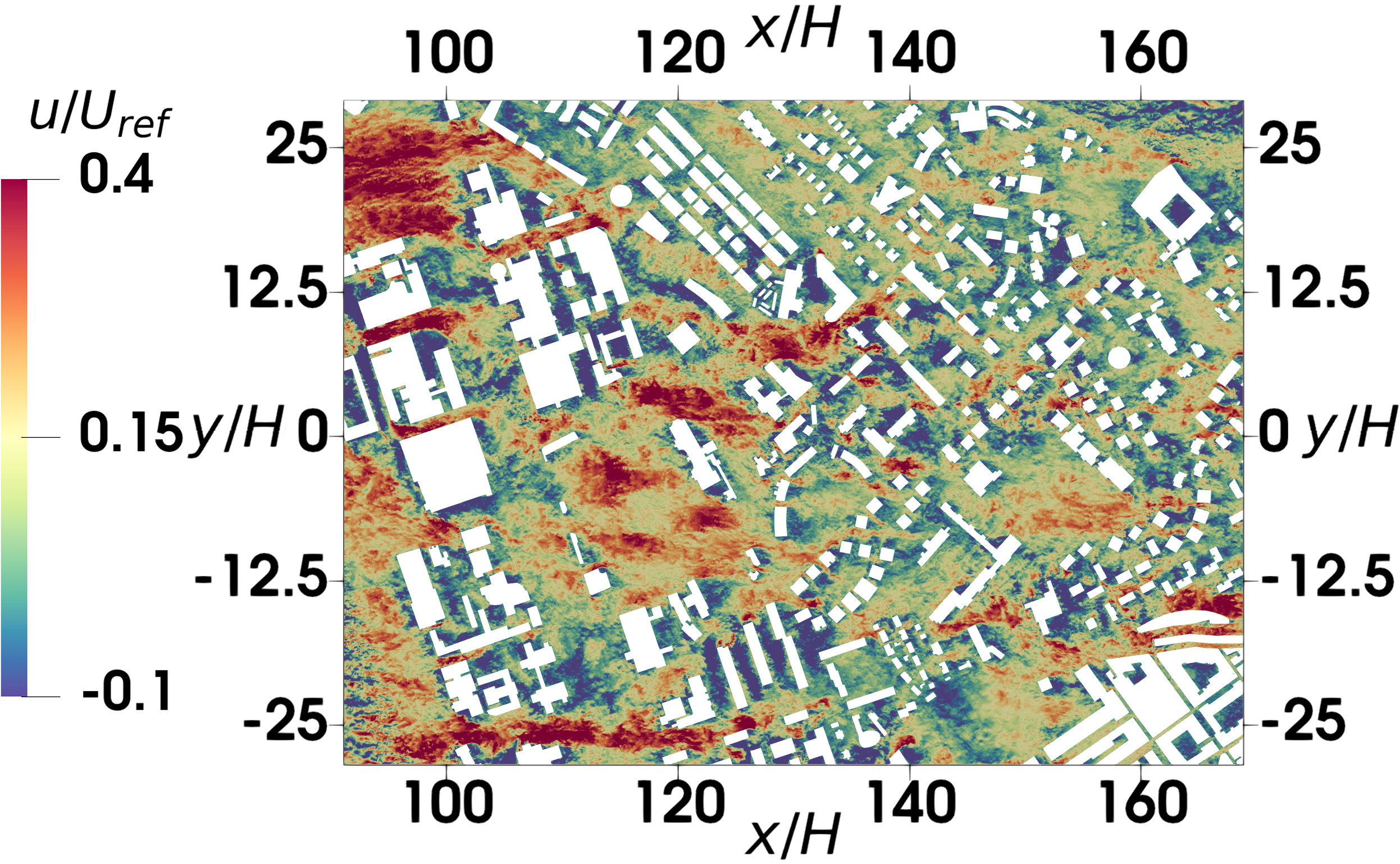}
    \caption{}
    \label{fig:subfig_a}
  \end{subfigure}
  \hfill
  \begin{subfigure}[b]{0.32\textwidth}
    \includegraphics[width=\textwidth]{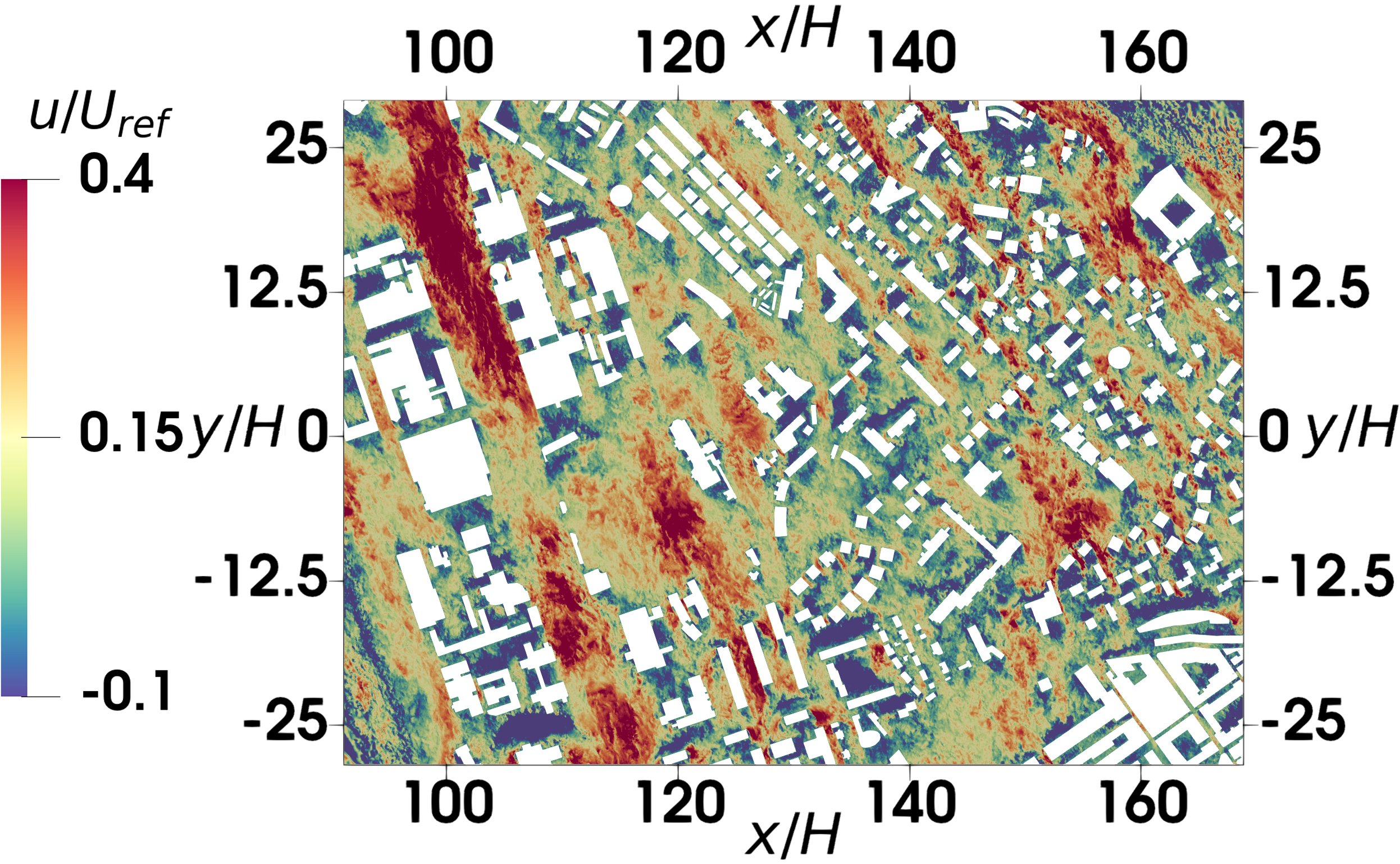}
    \caption{}
    \label{fig:subfig_b}
  \end{subfigure}
  \hfill
  \begin{subfigure}[b]{0.32\textwidth}
    \includegraphics[width=\textwidth]{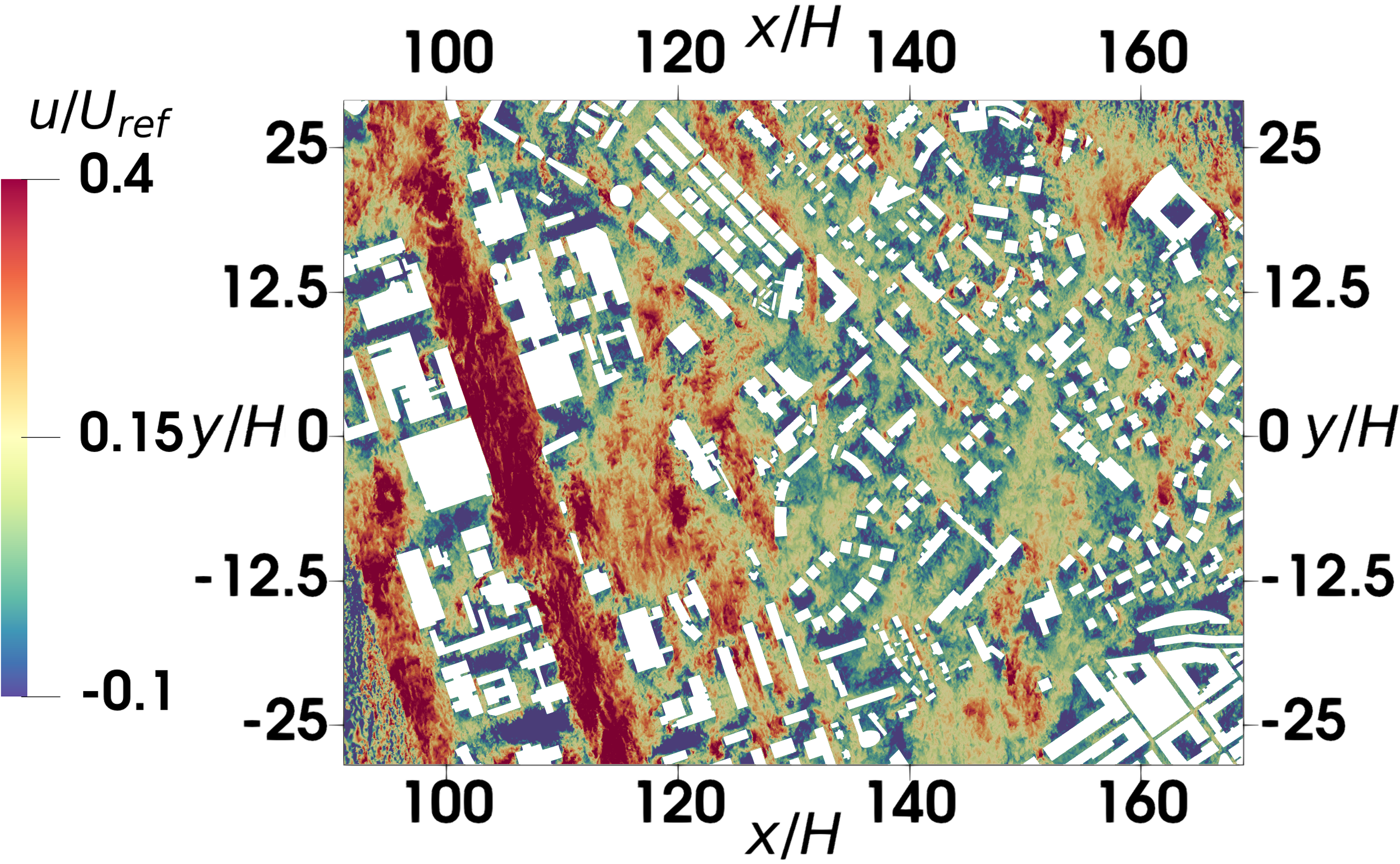}
    \caption{}
    \label{fig:subfig_c}
  \end{subfigure}
    \begin{subfigure}[b]{0.32\textwidth}
    \includegraphics[width=\textwidth]{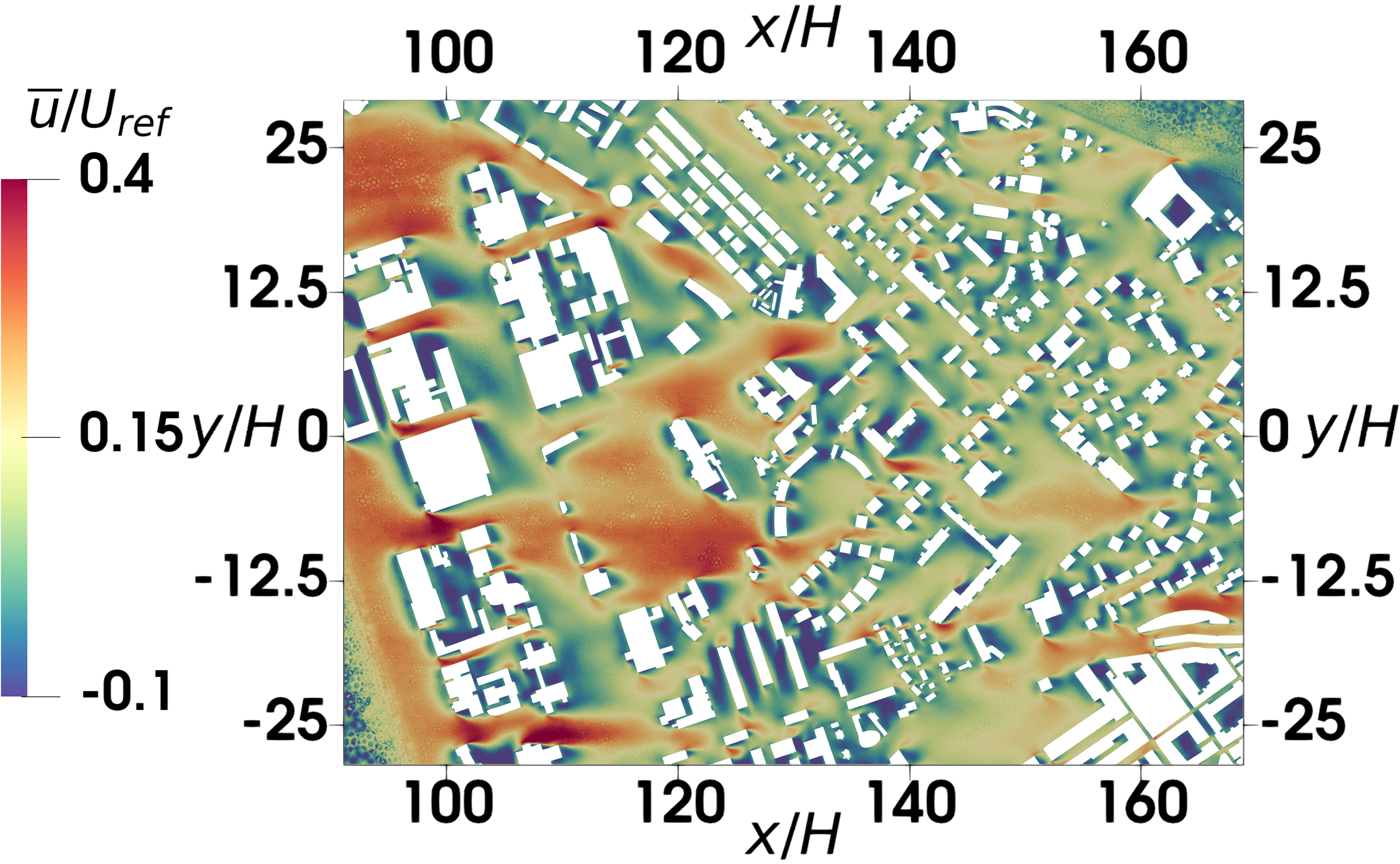}
    \caption{}
    \label{fig:subfig_a}
  \end{subfigure}
  \hfill
  \begin{subfigure}[b]{0.32\textwidth}
    \includegraphics[width=\textwidth]{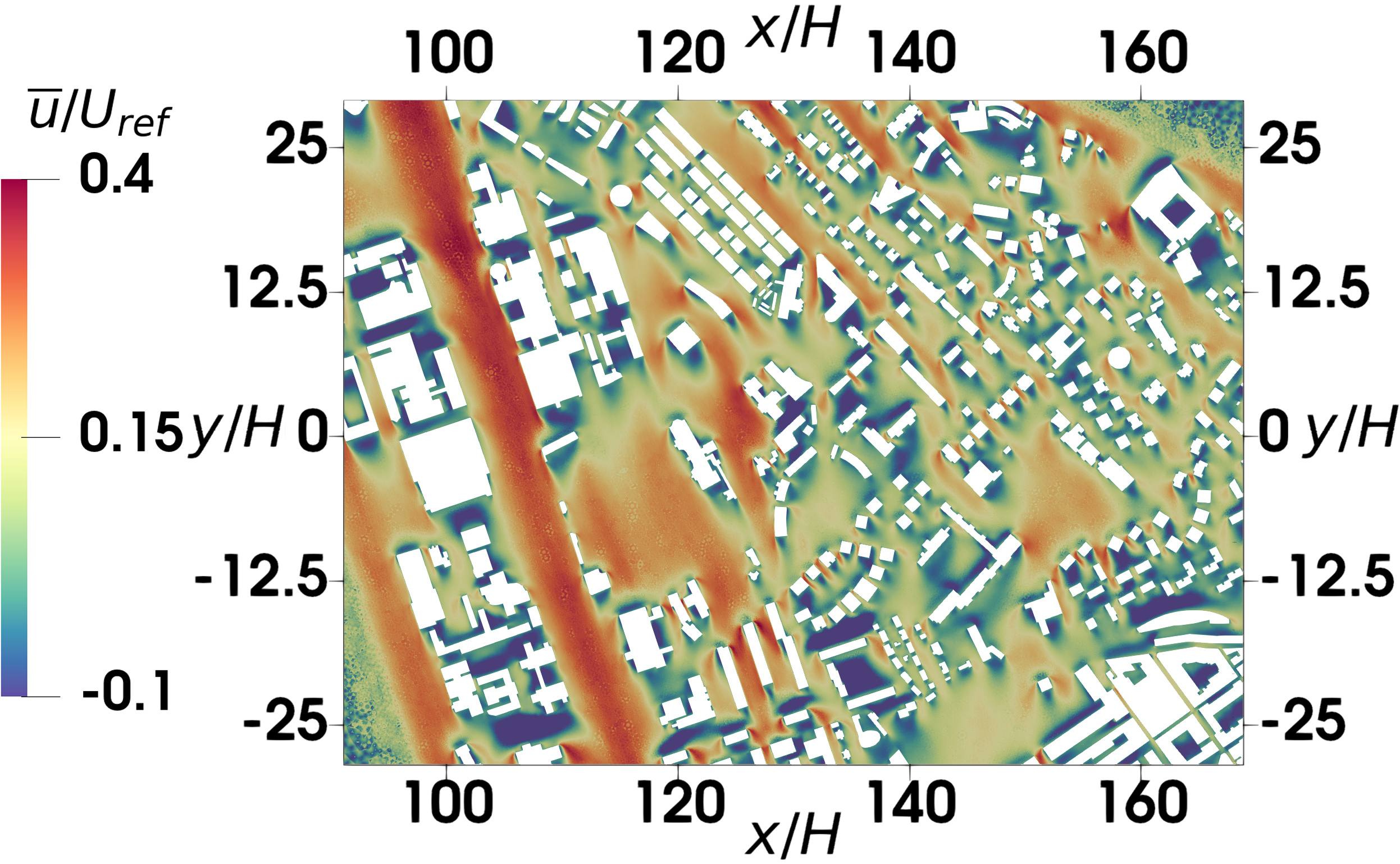}
    \caption{}
    \label{fig:subfig_b}
  \end{subfigure}
  \hfill
  \begin{subfigure}[b]{0.32\textwidth}
    \includegraphics[width=\textwidth]{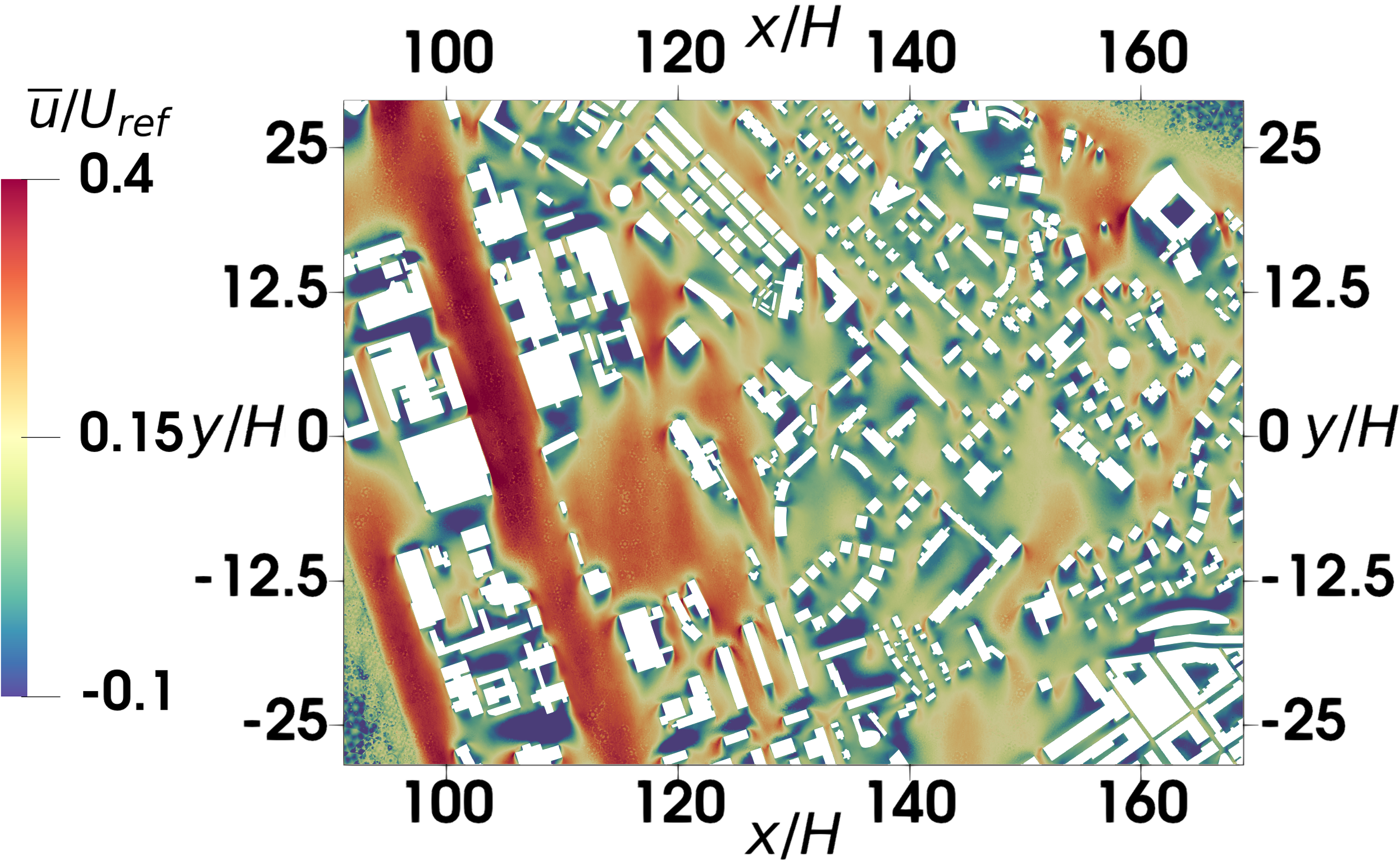}
    \caption{}
    \label{fig:subfig_c}
  \end{subfigure}
    \begin{subfigure}[b]{0.32\textwidth}
    \includegraphics[width=\textwidth]{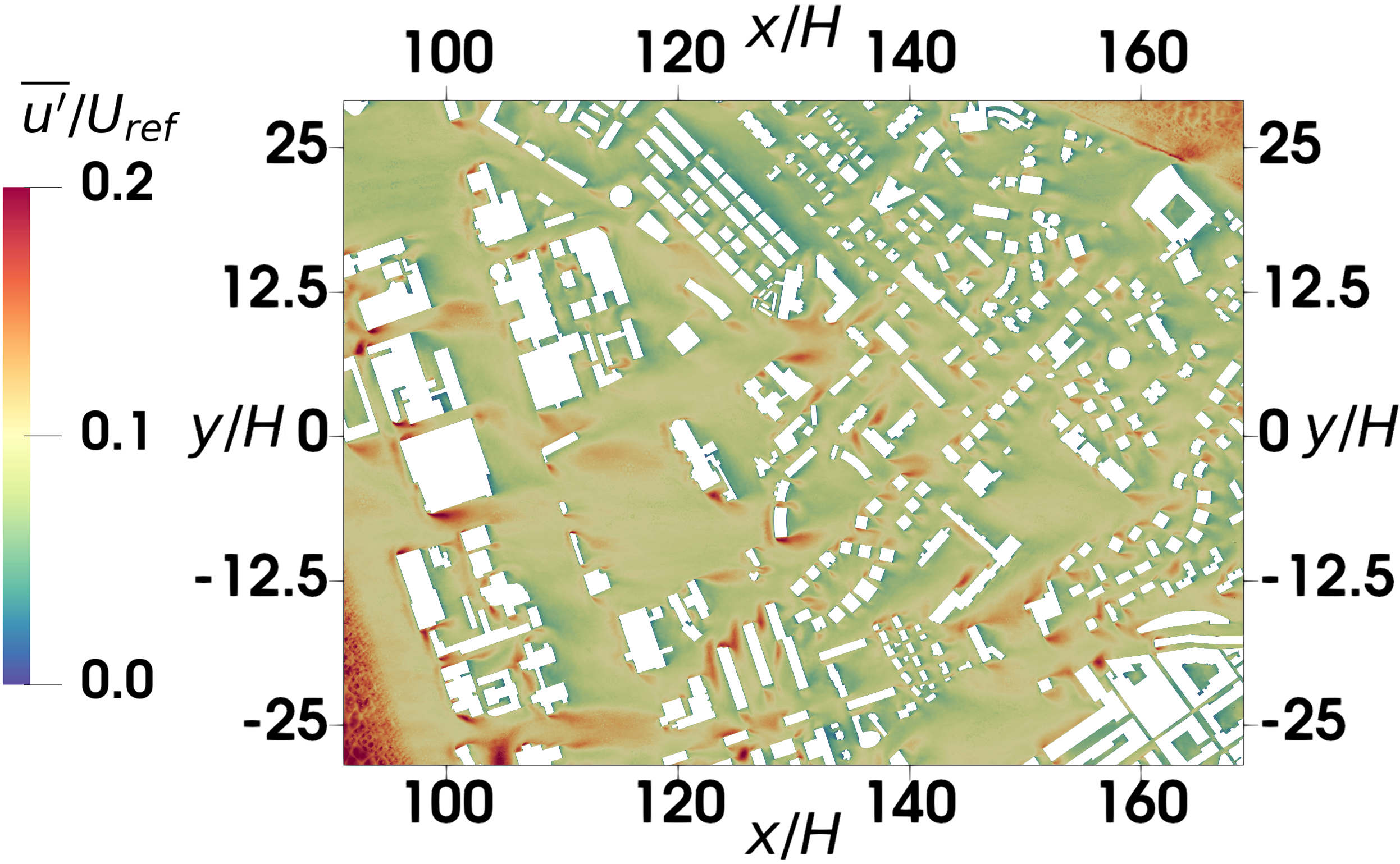}
    \caption{}
    \label{fig:subfig_a}
  \end{subfigure}
  \hfill
  \begin{subfigure}[b]{0.32\textwidth}
    \includegraphics[width=\textwidth]{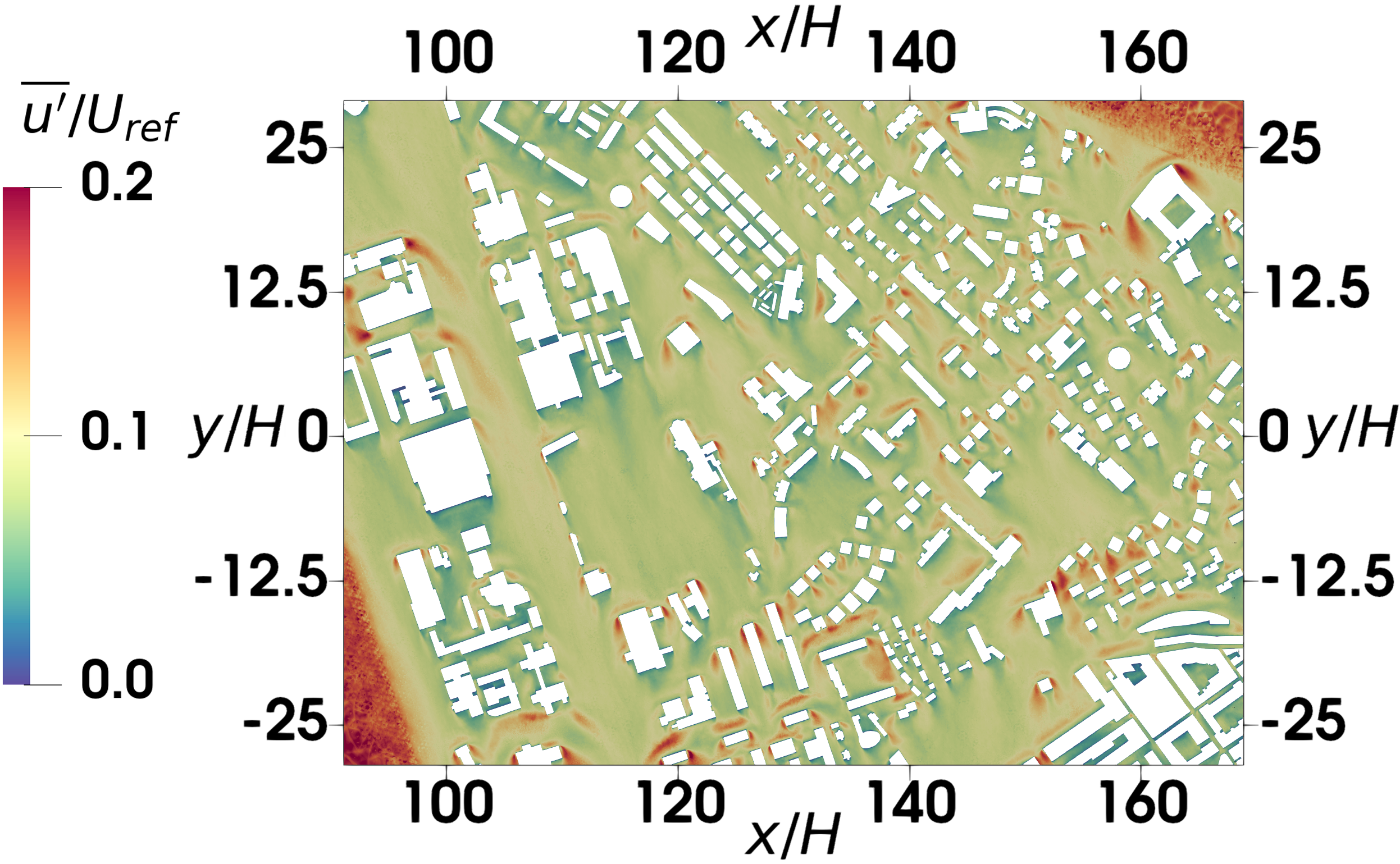}
    \caption{}
    \label{fig:subfig_b}
  \end{subfigure}
  \hfill
  \begin{subfigure}[b]{0.32\textwidth}
    \includegraphics[width=\textwidth]{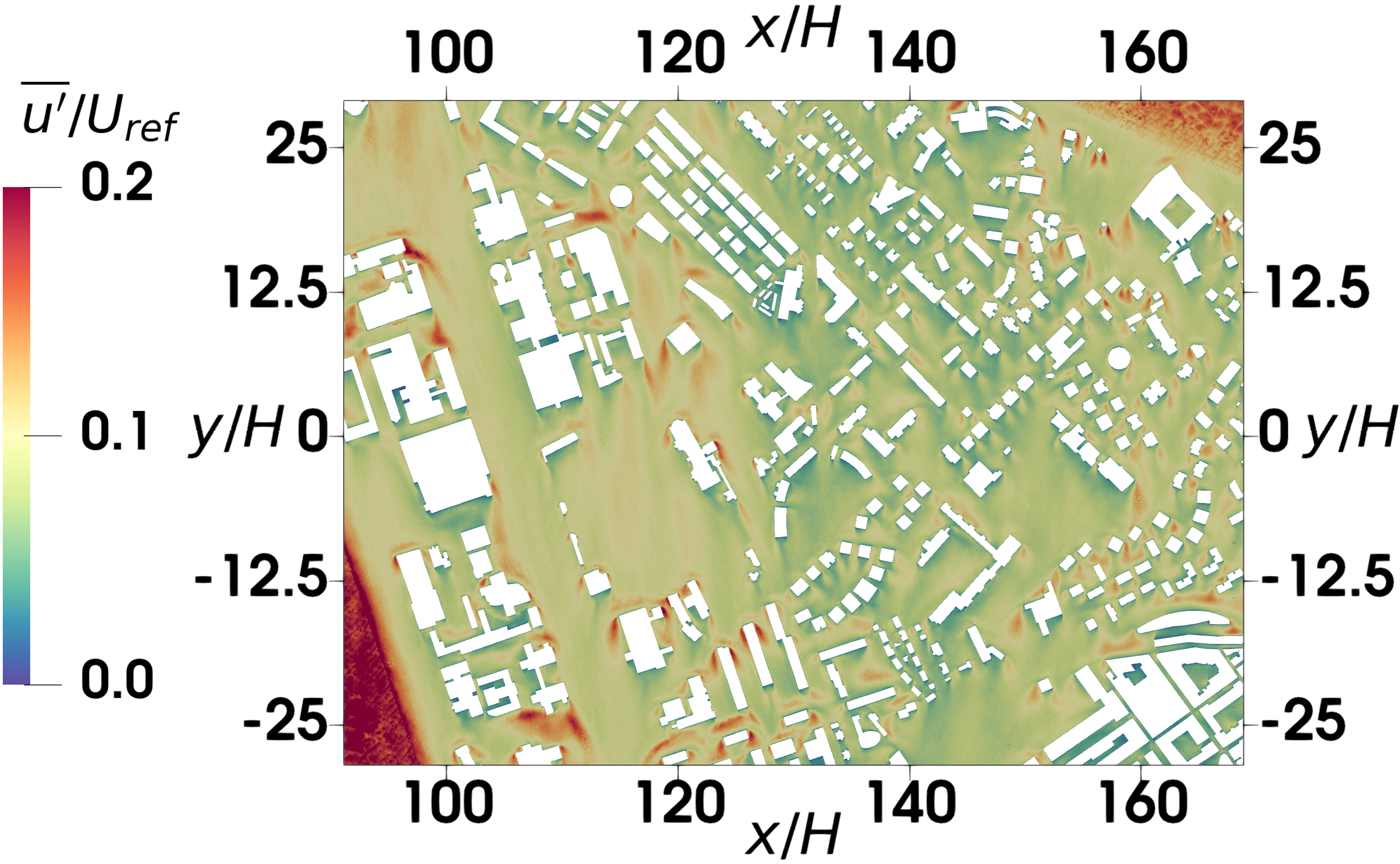}
    \caption{}
    \label{fig:subfig_c}
  \end{subfigure}
\caption{Effect of wind direction on pedestrian-level flow at 2 m height. Each column corresponds to a different wind direction ($0^\circ$, $67.5^\circ$, and $90^\circ$ from left to right). Panels (a–c) show instantaneous velocity fields, (d–f) the mean streamwise velocity, and (g–i) the turbulent kinetic energy. $U_{ref}$ is taken as the velocity at the top of the precursor domain.}
  \label{fig:pedestrian}
\end{figure*}

Figure \ref{fig:pedestrian} illustrates the impact of varying wind directions on the flow structure at a height of 2 meters, capturing instantaneous velocity (a–c), mean streamwise velocity (d–f), and turbulent kinetic energy (g–i). Across all cases, the complex morphology of the urban canopy induces significant spatial variability in the flow, including recirculation zones, channeling effects, and localized acceleration near building gaps.
At $0^\circ$ wind direction (left column), the flow enters the domain aligned with the main street orientation, producing well-developed wakes behind building blocks and extended regions of flow recirculation, especially in narrow street canyons. The mean velocity fields (panel d) show that large portions of the pedestrian layer remain sheltered from the main flow, with turbulent kinetic energy (panel g) concentrated near intersections and sharp edges.

When the inflow angle shifts to $67.5^\circ$ (middle column), the alignment between wind and street orientation becomes oblique, leading to greater flow penetration between buildings. The streamwise velocity (panel e) increases in various previously sheltered areas, while recirculating zones become more fragmented. This enhanced ventilation effect is also evident in the elevated levels of turbulent kinetic energy (panel h), particularly along diagonally aligned pathways.
Under perpendicular inflow at $90^\circ$(right column), strong channelling occurs along the dominant street axes, accelerating the flow in those directions and producing highly anisotropic patterns. The mean velocity field (panel f) shows continuous high-speed streaks along these corridors, whereas turbulent kinetic energy (panel i) is more confined to the shear zones near building edges. Compared to the $0^\circ$ case, the wakes behind buildings appear shorter and more rapidly recovered, likely due to enhanced mixing and turbulent diffusion introduced by cross-flow interaction.
In general, one can see that wind direction significantly modulates pedestrian-level flow. Oblique and perpendicular inflow configurations enhance the spatial heterogeneity of the urban canopy flow, reduce wake lengths, and increase turbulent kinetic energy in key regions.
\subsection{Flow statistics in the roughness sublayer}

\begin{figure}[t]
  \centering
  % Subfigura (a): Perfil u/u*
  \begin{subfigure}[b]{0.45\textwidth}
    \centering
    \includegraphics[width=\linewidth]{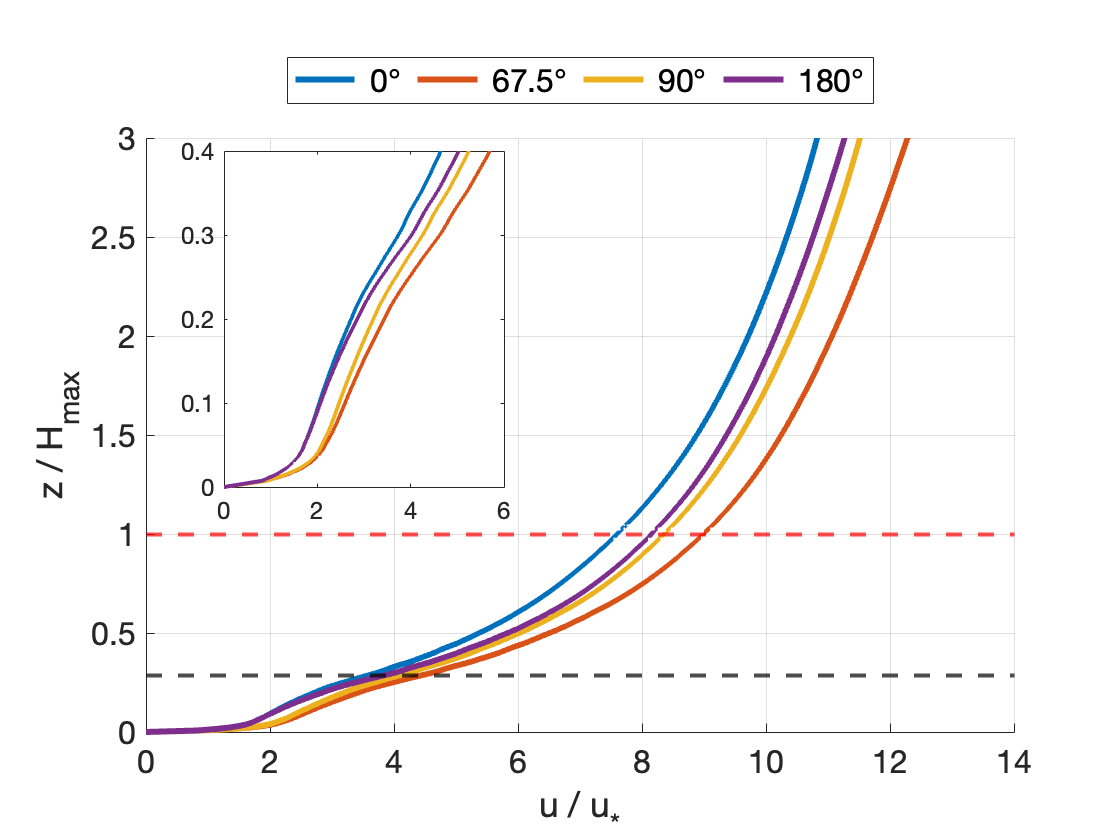}
    \caption{}
    \label{fig:uustar}
  \end{subfigure}
  \hfill
  % Subfigura (b): Perfil uu/u*^2
  \begin{subfigure}[b]{0.45\textwidth}
    \centering
    \includegraphics[width=\linewidth]{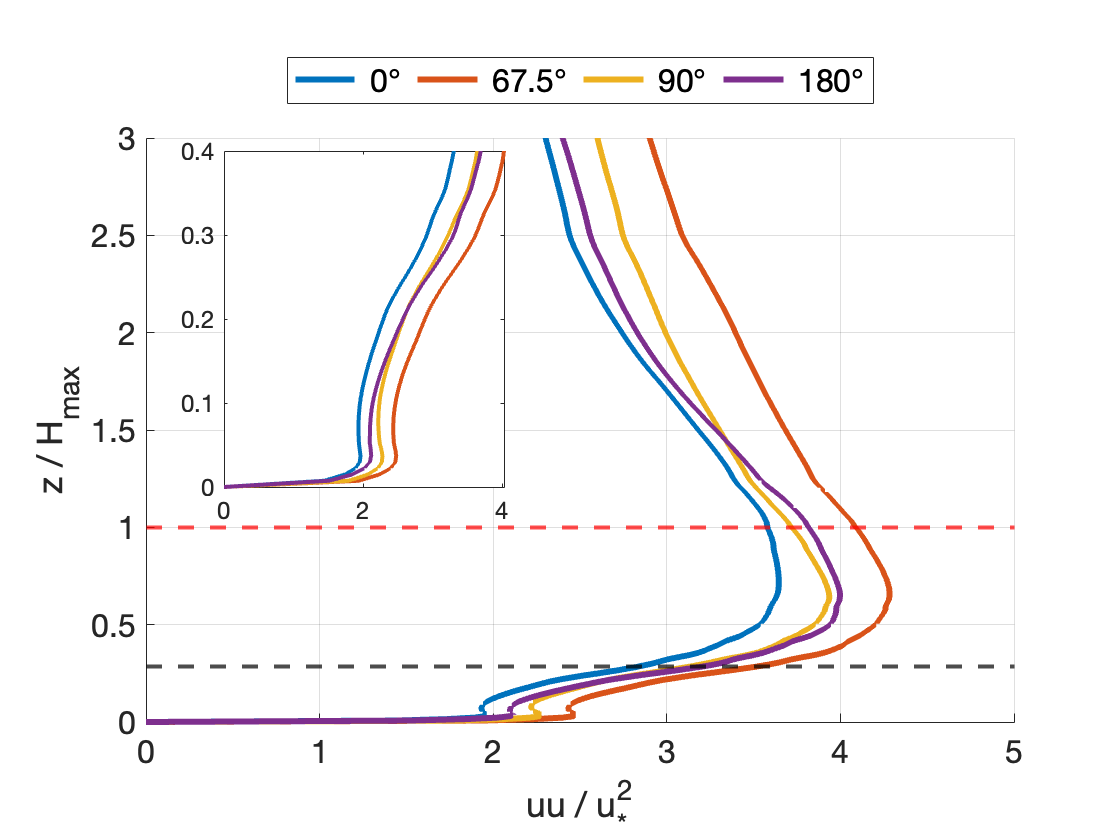}
    \caption{}
    \label{fig:uurms}
  \end{subfigure}
  \caption{Vertical profiles of (a) mean streamwise velocity and (b) streamwise velocity fluctuations, non-dimensionalized by the friction velocity $u_*$ for four wind directions ($0^\circ$, $67.5^\circ$, $90^\circ$, and $180^\circ$).}
  \label{fig:ustar_profiles}
\end{figure}

Figure~\ref{fig:ustar_profiles} presents the vertical profiles of the mean streamwise velocity $u/u_*$ and the streamwise velocity fluctuations $uu/u_*^2$, normalized by the local friction velocity $u_*$ for each wind direction ($0^\circ$, $67.5^\circ$, $90^\circ$, and $180^\circ$). Both quantities are shown as functions of the normalized height $z/H_{\text{max}}$, allowing for a direct comparison of the flow structure within and above the urban canopy layer.

A key feature common to all mean streamwise velocity profiles is the presence of an inflection point located below the average building height $H_{\text{avg}}$. This inflection point is generally interpreted as the effective height of the urban canopy layer, in line with the interpretation proposed by Oke et al.~(2017). The presence of such an inflection point suggests a mixing-layer-like flow regime, a behavior well documented in flows over vegetated canopies \cite{Raupach1996}. This supports the idea that, despite the geometric complexity of urban areas, the vertical exchange of momentum within the canopy is governed by mechanisms similar to those found in natural roughness layers.

In fact, this type of behavior has been observed in several previous studies. \cite{Giometto2016} reported that the inflection point in their velocity profiles was located near the average building height, $H_{\text{avg}}$. They attributed this  position to the presence of strong shear layers formed by the tallest buildings, which inhibit the downward penetration of turbulent structures originating from the flow above.
\cite{Cheng2023b} observed inflection points between $H_{\text{avg}}$ and $H_{\text{max}}$, which seems to be consistent with a more heterogeneous building height distribution of their study and stronger vertical shear.

 %These effects have direct implications for urban ventilation, pollutant dispersion, and thermal comfort assessments.

%Results from the study are shown in Fig. \ref{fig:F4} which showcases the capacity of the current methodology. In the figure, instantaneous maps of the velocity magnitude at pedestrian level are depicted for  $\phi=0^\circ$ and  $\phi=45^\circ$, together with the averaged flow. In the conference, the analysis of the effects of the incoming flow on the turbulence statistics will be presented and discussed in detail.
%To simulate the ABL in the wind tunnel experiments spires and floor roughness elements were used to obtain an equivalent ABL with reference velocity of 3m/s measured at building height H and friction velocity $u^*=0.24m/s$. 

In contrast, other studies such as \cite{akinlabi2022} report no clear inflection point in their velocity profiles. 
As they explain, the combination of uneven building shapes and varying heights, especially the presence of tall buildings and a large height standard deviation, generates intense vortical structures that interact strongly with the downstream environment. They especulate that this complex wake interaction enhances flow penetration into the canopy layer, disrupting the development of a clear inflection point and promoting high levels of mixing near the ground.

When comparing in more detail with \cite{Giometto2016} or cases 3 and 4 in \cite{Cheng2023b}, although our inflection point lies slightly below $H_{\text{avg}}$, there are clear similarities in terms of shape. These similarities stem from the more homogeneous building height distribution, with scarce very tall buildings, which would otherwise promote stronger shear layers and elevate the mixing layer top.

Interestingly, the influence of wind direction on the vertical profiles is relatively modest. All four directions ($0^\circ$, $67.5^\circ$, $90^\circ$, and $180^\circ$) yield similar velocity shapes, particularly below $z/H_{\text{max}} \approx 1$. This suggests that the effective aerodynamic roughness is not strongly directional in the studied area. A possible explanation is that the distribution of building orientations and heights is relatively isotropic or disordered, leading to a homogenized response when applying spatial and temporal averaging.

This observation is especially notable because instantaneous and temporally averaged flow fields do show significant directional effects, including acceleration within long street canyons (notably at $67.5^\circ$ and $90^\circ$), and varying degrees of stagnation or sheltering. However, these localized features tend to disappear in the double-averaged profiles, indicating that, at the scale of ensemble statistics, the urban canopy induces similar momentum transfer characteristics regardless of wind direction. This has important implications for urban flow modeling, suggesting that simplified or direction-agnostic parameterizations may be acceptable in areas lacking pronounced directional anisotropy in the built environment.

Unlike the inflection points identified in the mean velocity profiles, the peak in turbulent intensity occurs in between the average building height $H_{\text{avg}}$ and  the maximum  $H_{\text{max}}$ in all wind directions. This systematic shift in the location of peak turbulence intensity highlights the different physical mechanisms governing mean flow and turbulence generation. While the inflection point marks the top of the shear-driven mixing layer within the canopy, the $uu$ peak reflects regions of intense turbulent production and wake interaction.  The fact that the maximum  consistently occurs above the canopy layer suggests that turbulent stresses are not solely generated within the canopy, but are strongly influenced by advection and upward diffusion of wake structures originating on the roof of taller buildings. These elements introduce large-scale flow disturbances that penetrate above canopy layer. % Consequently, the position of the $uu$ peak appears to be dictated by the length scales set by the tallest obstacles, rather than the average canopy height. 

Moreover, the magnitude of the peak is significant in all cases, indicating robust and organized turbulent activity above the urban roughness elements. Compared to previous studies, these peaks occur at lower heights (below $H_{max}$), which can be attributed to the relatively homogeneous building height in the studied area and the limited presence of very tall structures. As a result, turbulence fluctuations appear to be governed by a combination of the production at average building height and the interaction with the few taller buildings. Nonetheless, these findings support the hypothesis that turbulent structures are primarily controlled by the dynamics occurring near rooftop levels, not only through the generation of strong shear layers but also by sheltering smaller elements and concentrating wake production above the urban canopy layer.

%This supports the idea that tall buildings dominate the turbulent structure \cite{Cheng2023b}, not only by generating strong shear but also by sheltering smaller elements and concentrating wake production at higher elevations. The consistent peak location near $H_{\text{max}}$ reinforces the notion that, in heterogeneous urban configurations, turbulence is regulated more by the tallest elements than by statistical metrics like $H_{\text{avg}}$.

\section{Conclusions}
The present study has demonstrated that, even under varying wind directions, the overall structure of the flow within a dense urban canopy remains robust when analyzed through double averaged statistics. This suggests that, beyond the immediate complexity of local flow features, the urban morphology imposes dominant constraints that organize the vertical distribution of momentum and turbulence.
The location of the velocity inflection point, consistently found below the average building height, confirms the presence of a mixing layer type regime within the canopy. In contrast, turbulence intensity peaks occur closer to the tallest buildings, highlighting their role as key turbulence generating elements. Interestingly, the limited directional dependence observed in the profiles implies that, at the scale of urban blocks, aerodynamic roughness behaves quasi isotropically. This has implications for modeling strategies, indicating that direction agnostic formulations may be suitable for certain urban canopy applications.
Furthermore, the simulations presented here contribute to the creation of a high fidelity database over a realistic urban geometry. This dataset is intended to support the validation of urban canopy models and the development of surrogate models for fast and reliable flow prediction in complex city environments.
\section*{Acknowledgments}
%\vspace{-0.5cm}
This work has been partially financially supported by 'Agència de Gestió d'Ajuts Universitaris i de Recerca' under the call CLIMA 2023 (ref. 2023 CLIMA 00097). The authors acknowledge the support
of the Departament de Recerca i Universitats de la Generalitat de Catalunya through the research
group Large-scale Computational Fluid Dynamics (ref.: 2021 SGR 00902) and the Turbulence and
Aerodynamics Research Group (ref.: 2021 SGR 01051). 
The authors also acknowledge Red Española de Supercomputacion the resources provided in Marenostrum V Supercomputer (IM-2024-2-0006, IM-2024-3-0006).
%We also acknowledge Red Espa\~nola de Surpercomputaci\'on (RES) for awarding us access
%to the MareNostrum IV machine based in Barcelona, Spain (Ref IM-2024-3-0006).
 
%%% References %%%
\bibliographystyle{agsm}
%\bibliography{CitiesBib}
%
%\begin{References}
%\item Jones A.B., Jordan, D.L and March, F.P. (2002), Linear aspects of transition, {\it J. Fluid Mech.}, Vol. 34, pp. 123-153. 
%\item Smith, W.S. and Martin, W.P. (1997), Leading edge paper on control mechanism, AIAA Paper 97-1234.
%\end{References}

\end{document}